\documentclass[english,aps,prd,floats,floatfix,nofootinbib]{revtex4}
\usepackage{rotating}
\usepackage{natbib}
\usepackage{times}
\usepackage{float}
\usepackage{amsmath}
\usepackage{graphicx}
\usepackage{amssymb}
\usepackage{slashed}
\usepackage{color}
\usepackage{tikz}
\usetikzlibrary{shapes,arrows}

\tikzstyle{decision} = [diamond, draw, fill=blue!20, text width=4.5em, text badly centered, node distance=2.5cm, inner sep=0pt]
\tikzstyle{block} = [rectangle, draw, fill=blue!20, text width=6em, text centered, rounded corners, minimum height=4em]
\tikzstyle{wideblock} = [rectangle, draw, fill=blue!20, text width=8em, text centered, rounded corners, minimum height=4em]
\tikzstyle{redblock} = [rectangle, draw, fill=red!20, text width=5em, text centered, rounded corners, minimum height=4em]
\tikzstyle{yellowblock} = [rectangle, draw, fill=yellow!20, text width=5em, text centered, rounded corners, minimum height=4em]
\tikzstyle{wideyellowblock} = [rectangle, draw, fill=yellow!20, text width=8em, text centered, rounded corners, minimum height=4em]
\tikzstyle{line} = [draw, very thick, color=black!50, -latex']
\tikzstyle{dashedline} = [draw, very thick, color=black!50, -latex',dashed]
\tikzstyle{cloud} = [draw, ellipse,fill=red!20, node distance=2.5cm, minimum height=2em]
\tikzstyle{decision answer}=[draw, ellipse,fill=green!20, node distance=2.5cm, minimum height=2em]

\newcommand{\eL}{\epsilon_L}
\newcommand{\eR}{\epsilon_R}
\newcommand{\eS}{\epsilon_S}
\newcommand{\eP}{\epsilon_P}
\newcommand{\eT}{\epsilon_T}

\newcommand{\beq}{\begin{equation}} 
\newcommand{\eeq}{\end{equation}} 
\newcommand{\ba}{\begin{array}}  
\newcommand{\ea}{\end{array}} 
\newcommand{\bea}{\begin{eqnarray}}  
\newcommand{\eea}{\end{eqnarray} }  
\newcommand{\be}{\begin{eqnarray}}  
\newcommand{\ee}{\end{eqnarray} }  
\newcommand{\bal}{\begin{align}}
\newcommand{\eal}{\end{align}}   
\newcommand{\bi}{\begin{itemize}}  
\newcommand{\ei}{\end{itemize}}  
\newcommand{\ben}{\begin{enumerate}}  
\newcommand{\een}{\end{enumerate}}  
\newcommand{\bc}{\begin{center}}
\newcommand{\ec}{\end{center}} 
\newcommand{\bt}{\begin{table}}
\newcommand{\et}{\end{table}}  
\newcommand{\btb}{\begin{tabular}}
\newcommand{\etb}{\end{tabular}}

\usepackage[normalem]{ulem}
\usepackage[makeroom]{cancel}

\begin{document}
\title{New Physics in $s\to u\ell^-\bar\nu$: Interplay between semileptonic kaon and hyperon decays}

\author{Mart\'{i}n Gonz\'{a}lez-Alonso}%
\affiliation{%
IPN de Lyon/CNRS, Universite Lyon 1, Villeurbanne, France
}
\author{Jorge Martin Camalich}
\affiliation{PRISMA Cluster of Excellence Institut f\"ur Kernphysik, 
Johannes Gutenberg-Universit\"at Mainz, 55128 Mainz, Germany}

\begin{abstract}
We review a novel model-independent approach to the analysis of new-physics effects in the $s\to u\ell^-\bar\nu$ transitions. 
We apply it to (semi)leptonic kaon decays and study their complementarity with pion and hyperon $\beta$ decays or with collider searches of 
new physics.
\end{abstract}
\maketitle

\section{Introduction}

The $K\to\pi\ell\nu$ ($K_{\ell3}$) and $P\to \ell\nu$ ($P_{\ell2}$) decays, where $P=\pi,K$ and $\ell=e,\,\mu$, boast one of the most precise data bases
in hadronic weak decays~\cite{Gonzalez-Alonso:2016etj,Antonelli:2008jg,Antonelli:2010yf,Cirigliano:2011ny,Agashe:2014kda}. The hadronic form factors necessary to describe these processes
are flagship quantities for lattice QCD (LQCD) and the theoretical accuracy at which they are calculated is now below the percent level~\cite{Aoki:2013ldr,Rosner:2015wva}. 
Much work has also been done in Chiral Perturbation Theory (ChPT) and using dispersive methods to understand analytically low-energy theorems and small 
contributions to the decay rates such as isospin breaking and the electromagnetic radiative corrections~\cite{Gasser:1984gg,Gasser:1984ux,Cirigliano:2001mk,
Bijnens:2003uy,Jamin:2001zq,Ananthanarayan:2004qk,Jamin:2004re,Cirigliano:2005xn,DescotesGenon:2005pw,Cirigliano:2007xi,Cirigliano:2007ga,Kastner:2008ch,
Cirigliano:2008wn,Bernard:2009zm,Cirigliano:2011tm}. This is of great interest at the moment given the various hints of New Physics (NP) spotted in flavor 
observables~\cite{Lees:2012xj,Aaij:2013qta,Aaij:2014ora,Aaij:2015yra,Huschle:2015rga} and at the LHC~\cite{Aaboud:2016tru,Khachatryan:2016hje}. 

Effective Field Theories (EFT) are optimal tools to perform such analysis, and to test the Standard Model (SM) in a model-independent way. 
Although there is work on bounds to specific scenarios of NP~\cite{Bernard:2006gy} and the EFT language is introduced in 
the Flavianet analyses~\cite{Antonelli:2008jg,Antonelli:2010yf}, an EFT approach has not been used for global studies 
of the $s\to u$ transitions yet beyond CKM unitarity tests by precise determinations of $|V_{us}|$ and $|V_{ud}|$~\cite{Cirigliano:2009wk,Antonelli:2010yf,Aoki:2013ldr}. 
Notice the difference with the $d\to u$ decays, where global EFT fits have been performed 
by various groups~\cite{Severijns:2006dr,Wauters:2013loa,Pattie:2013gka}. 

In the following, we briefly review the extension of this program to 
the $s\to u\ell\nu$ transitions. For a comprehensive discussion and a more complete list of references we refer the reader to Ref.~\cite{Gonzalez-Alonso:2016etj}.

\section{Effective Lagrangians}

The starting point is the low-scale  $O(1 \ {\rm GeV})$  effective Lagrangian for semi-leptonic $D \to u$ transitions ($D=s,\,d$)~\cite{Cirigliano:2009wk}:
\begin{eqnarray}
{\cal L}_{\rm eff} 
&=&
- \frac{G_F^{(0)} V_{uD}}{\sqrt{2}}\,\sum_{\ell=e,\mu}
\Bigg[
\Big(1 + \epsilon_L^{D \ell}  \Big) \bar{\ell}  \gamma_\mu  (1 - \gamma_5)   \nu_{\ell} \cdot \bar{u}   \gamma^\mu (1 - \gamma_5 ) D
+  \eR^{D\ell}  \   \bar{\ell} \gamma_\mu (1 - \gamma_5)  \nu_\ell    \ \bar{u} \gamma^\mu (1 + \gamma_5) D\nonumber\\
&+& \bar{\ell}  (1 - \gamma_5) \nu_{\ell}
\cdot \bar{u}  \Big[  \epsilon_S^{D\ell}  -   \epsilon_P^{D\ell} \gamma_5 \Big]  D
+ \epsilon_T^{D\ell}    \,   \bar{\ell}   \sigma_{\mu \nu} (1 - \gamma_5) \nu_{\ell}    \cdot  \bar{u}   \sigma^{\mu \nu} (1 - \gamma_5) D
\Bigg]+{\rm h.c.}, 
\label{eq:leff1} 
\end{eqnarray}
where $G_F^{(0)}\equiv \sqrt{2}g^2/(8 M_W^2)$ and the $\epsilon_i^{D\ell}$ are Wilson coefficients (WC) which carry a $\sim v^2/\Lambda^2$ dependence
on the NP scale $\Lambda$. The effective operators undergo renormalization under quantum corrections in the SM and the WC can display
renormalization scale dependence, which is canceled by the opposite one of the hadronic matrix elements at the level of the 
observables (see~\cite{Gonzalez-Alonso:2016etj} for details).

The EFT Lagrangian in Eq.~(\ref{eq:leff1}) mediates all low-energy charged-current weak processes involving up and generic down-type 
quarks like pions, kaons, neutrons and hyperons. In total, the EFT contains 20 WC describing  model-independently the most general NP modifications
to the charged-current decays $D\to u\ell\nu$. 

If the NP is coming from dynamics at $\Lambda\gg v$ 
and EWSB is linearly realized, then one can use an $SU(2)_L\times U(1)_Y$ EFT~\cite{Buchmuller:1985jz,Cirigliano:2009wk}. A non-trivial
consequence of this is that, at leading order in the matching between the high- and low-energy theories:
\begin{equation}
\epsilon_R^{De}=\epsilon_R^{D\mu}+\mathcal{O}(v^4/\Lambda^4)\equiv \epsilon_R^{D}.\label{eq:RHCuniversal} 
\end{equation}
We will neglect $\mathcal{O}(v^4/\Lambda^4)$ throughout this note so that a NP effect involving
a right-handed current necessarily involves a Higgs-current fermion-current operator~\cite{Buchmuller:1985jz} and its contribution must 
be lepton universal. In addition, the number of independent WC in the EFT gets reduced from 20 to 18. One can see that the CKM matrix 
element gets a NP contribution such that one can only extract the following combination:
\begin{equation}
\tilde V_{uD}^\ell=\left(1 + \epsilon_L^{D\ell} + \epsilon_R^{D}-\frac{\delta G_F}{G_F}\right)\,V_{uD}+\mathcal{O}(v^4/\Lambda^4).\label{eq:VuDNPs}
\end{equation}
where $\delta G_F$ contains the NP effect affecting the experimental extraction of the Fermi constant from the muon lifetime~\cite{Cirigliano:2009wk}.
Consequently, these NP terms can only be probed through CKM unitarity tests or through lepton-universality tests.

\section{Decay observables}

\subsection{$\pi_{\ell2(\gamma)}$ and $K_{\ell2(\gamma)}$}

In order to minimize various experimental and theoretical uncertainties, it is convenient to work with the three ratios 
$R_P=\Gamma_{P_{e2}(\gamma)}/\Gamma_{P_{\mu2}(\gamma)}$  and $R_\mu=\Gamma_{K_{\mu2}(\gamma)}/\Gamma_{\pi_{\mu2}(\gamma)}$, which are given by
\begin{align}
R_P
&\equiv \frac{n_{Pe}}{n_{P\mu}} \frac{|\tilde V^e_{uD}|^2}{|\tilde V^\mu_{uD}|^2}\left(1+\Delta^P_{e2/\mu2}\right)\label{eq:epsDemu}\\
&= \frac{n_{Pe}}{n_{P\mu}} \left( 1+2\left(\eL^{De}-\eL^{D\mu}\right)-2B_0\left(\frac{\eP^{De}}{m_e}-\frac{\eP^{D\mu}}{m_\mu}\right)\right) +\mathcal{O}(v^4/\Lambda^4),\label{eq:deltaRPexp2}\\
R_\mu 
&= \frac{n_{K\mu}}{n_{\pi\mu}} \frac{|\tilde V_{us}^\ell|^2\,f_{K^\pm}^2}{|\tilde V_{ud}^\ell|^2\,f_{\pi^\pm}^2}\left(1+\Delta^{K/\pi}_{\ell2} \right)
\label{eq:epslKpi}\\
&= \frac{n_{K\mu}}{n_{\pi\mu}}\frac{|\tilde V_{us}^\ell|^2\,f_{K^\pm}^2}{|\tilde V_{ud}^\ell|^2\,f_{\pi^\pm}^2}\left(1-4(\eR^s-\eR^d)-
\frac{2B_0}{m_\ell}\left(\eP^{s\ell}-\eP^{d\ell}\right)\right)+\mathcal{O}(v^4/\Lambda^4).
\label{eq:deltarl1}
\end{align}
where $D=d,\,s$ for $P=\pi,\,K$ respectively, $f_{P^{\pm}}$ is the corresponding QCD semileptonic decay constant~\cite{Aoki:2013ldr}, 
$n_{P\ell}= m_P\,m_\ell^2(1- m_\ell^2/m_P^2)^2 (1+ \delta^{P\ell}_{\rm em})$, and $\delta^{P\ell}_{\rm em}$ is the corresponding electromagnetic correction~\cite{Cirigliano:2007ga}. We have used $m_P^2/(m_u+m_D)\simeq\, B_0$, and we denote by $\Delta$ the NP correction not absorbed in $\tilde V_{uD}^\ell$. We also need to include in the analysis one total rate (controlling the overall normalization) that we choose to be that of $K_{\mu2(\gamma)}$:
\begin{align}
\Gamma_{K_{\mu2}(\gamma)} 
&=\frac{G_F^2|\tilde V_{us}^\mu|^2\,f_{K^{\pm}}^2}{8\pi} n_{K\mu} \left(1+\Delta^K_{\mu2}\right)
\nonumber\\
&=\frac{G_F^2|\tilde V_{us}^\mu|^2\,f_{K^{\pm}}^2}{8\pi} n_{K\mu} \left(1-4\eR^s -\frac{2B_0}{m_\mu}\eP^{s\mu}\right)+\mathcal{O}(v^4/\Lambda^4). \label{eq:Kmu2}
\end{align}

\subsection{$K_{\ell3(\gamma)}$}

The $K\to\pi\ell\nu$ decay amplitude depends on the hadronic form factors $f_+(q^2)$, $f_0(q^2)$ and $B_T(q^2)$, which are defined following 
the conventions in Refs.~\cite{Antonelli:2008jg,Antonelli:2010yf,Gonzalez-Alonso:2016etj}. The photon-inclusive $K_{\ell3}$ total decay 
rates can be written as~\cite{Gonzalez-Alonso:2016etj}
\begin{equation}
\Gamma(K_{\ell 3(\gamma)}) =
{G_F^2 m_K^5 \over 192 \pi^3}\, C\,
  S_{\rm ew}\,|\tilde{V}_{us}^{\ell}|^2 f_+(0)^2\,
I_K^\ell(\lambda_{+,0},\,\epsilon^{s\ell}_{S,T})\,\left(1 + \delta^c+\delta^{c\ell}_{\rm
em}\right)^2\,\label{eq:Kl3rate},
\end{equation}
where $C=1$ ($1/2$) for the neutral (charged) kaons, $\delta^{c\ell}_{\rm em}$ are
radiative corrections and $\delta^c$ is the isospin-breaking correction for the charged kaon channel. 
The $I_K^\ell(\lambda_{+,0},\epsilon^{s\ell}_{S,T})$ is the phase space integral that depends on the scalar
and tensor WC and the so-called slopes, which parametrize the $q^2$ dependence of the form factors and that we have symbolically denoted here by $\lambda_{+,0}$. 
The specific expression of the phase space integral can be found in Ref.~\cite{Gonzalez-Alonso:2016etj}. 

It is interesting to note that for $f_0(q^2)$ one can use a dispersive representation \cite{Jamin:2001zq,Bernard:2006gy,Bernard:2009zm} with a single measurable free parameter 
that is conveniently chosen to be $C=f_0(m_K^2-m_\pi^2)/f_+(0)$, because its value can be determined in QCD using the Callan-Treiman theorem (CTT)~\cite{Callan:1966hu}
\begin{equation}
C_{\scriptscriptstyle{\rm QCD}}=\frac{f_0(m_K^2-m_\pi^2)}{f_+(0)}=\frac{f_K}{f_\pi}\frac{1}{f_+(0)}+\mathcal{O}(m_{u,d}/(4\pi f_\pi)),\label{eq:CTth}
\end{equation}
where the corrections have been calculated in ChPT at the level of $10^{-3}$~\cite{Gasser:1984ux}.

These QCD parameters ($C$ and $\lambda_+$) are customarily fitted to the kinematic distributions of the decay (or Dalitz plot), enabling a calculation 
the phase-space integral under the assumption $\epsilon_{S}^{s\ell}=\epsilon_{T}^{s\ell}=0$. However, scalar and tensor operators modify the phase space integrals and they should be determined together with the 
form factor parameters in the fits to the kinematic distributions. 
Importantly, their interference with the SM is helicity suppressed and enters proportional to the lepton mass, so that 
the kinematic distributions for the electronic mode are not sensitive to scalar and tensor operators at leading order of the EFT expansion~\footnote{Nonetheless, 
bounds can be obtained extending the analysis to include these quadratic terms~\cite{Gonzalez-Alonso:2016etj}.}.

In the muon channel, the effect of the scalar operator is absorbed in the $q^2$ dependence of the scalar form factor. 
If precise values for $f_+(0)$ and $f_K/f_\pi$ are provided in QCD, then the CTT gives a very accurate prediction of this 
form factor at $q^2=m_K^2-m_\pi^2$, which allows one to separate $\epsilon_S^{s\mu}$ from the experimental measurement:
\begin{align}
\log C_{\rm exp}
&=\log \left[C_{\scriptscriptstyle{\rm QCD}} \left( 1+ \frac{m_K^2-m_\pi^2}{m_\mu(m_s-m_u)} \epsilon_S^{s\mu} \right)\right],\label{eq:lnCtilde}
\end{align}
where we have used $\log C$ as suggested in~\cite{Bernard:2006gy} and adopted by the experimental collaborations~\cite{Antonelli:2010yf}.  

The tensor term can not be described by a simple re-definition of the SM contributions. This is clearly visible in the two-fold differential 
decay rate~\cite{Gonzalez-Alonso:2016etj} and a bound on tensor operators must be obtained from a global fit to the Dalitz plot of the form-factor
parameters and the WC $\epsilon_T^{s\ell}$, provided we have a LQCD result for the tensor form factor $B_T(0)$~\cite{Baum:2011rm}. Nonetheless, in order to assess the overall sensitivity of the muonic channel to the tensor operator, let us separate $I_K^\mu(\lambda_{+,0},
\,\epsilon^{s\mu}_{S,T})=I_{K,0}^\mu-\eT^{s\mu}I_T^\mu+\mathcal{O}(v^4/\Lambda^4)$, where $I_T^\mu/I_{K,0}^\mu\sim 0.40$ with the numerical inputs of Ref.~\cite{Gonzalez-Alonso:2016etj}. This will be of interest when comparing with the sensitivity of the hyperon $\beta$ decays at the level of the total rates.

Finally, we notice that the form factor $f_+(0)$ cancels in the lepton-universality ratio
\begin{align}
r_{\mu e} &=\frac{\Gamma_{K_{\mu3}} I_{e3} \left(1+2\delta_{\rm em}^{Ke}\right) }{\Gamma_{K_{e3}} I_{\mu3} \left(1+2\delta_{\rm em}^{K\mu}\right)} 
={\frac{|\tilde V_{us}^\mu|^2}{|\tilde V_{us}^e|^2} =
1+2\Delta^s_L}+\mathcal{O}(v^4/\Lambda^4)
~,\label{eq:Kl3LUrate}
\end{align}
that is only sensitive to the difference of left-handed $s\to u$ currents, $\Delta^s_L = \eL^{s\mu}-\eL^{se}$, up to $\mathcal{O}(v^2/\Lambda^2)$ due to Eq.~(\ref{eq:RHCuniversal}).

\subsection{Nuclear, neutron and hyperon $\beta$ decay}

The study of superallowed nuclear $\beta$ transitions provide the most accurate determination of $|\tilde V_{ud}^e|$ along with the most stringent limits on the scalar Wilson coefficient $\epsilon_S^{de}$~\cite{Hardy:2014qxa,Gonzalez-Alonso:2013uqa}.
Combining this $|\tilde V_{ud}^e|$ determination with the $|\tilde V_{us}^{e}|$ value obtained from $K_{e3(\gamma)}$ allows 
one to test CKM unitarity, which probes the following combination of WC~\cite{Cirigliano:2009wk}:
\begin{align}
\Delta_{\rm CKM}=1- |\tilde V_{ud}^e|^2+|\tilde V_{us}^e|^2=2|\tilde V_{ud}^e|^2(\eL^{de}+\eR^{d})+2|\tilde V_{us}^e|^2&(\eL^{se}+\eR^{s})-2\frac{\delta G_F}{G_F}+\mathcal{O}(v^4/\Lambda^4).\label{eq:CKMuni}
\end{align}
where we have neglected the contribution of $|\tilde{V}_{ub}^{e}|^2$ because its value is smaller than the current uncertainty in $\Delta_{\rm CKM}$~\cite{Agashe:2014kda}. 

Neutron and hyperon $\beta$ decays are sensitive to both vector and axial currents, which come weighed by different form 
factors~\cite{Weinberg:1958ut,Bhattacharya:2011qm,Chang:2014iba}. While one obtains a value for $|\tilde V_{uD}^\ell|$ once the ``vector charge'' of the transition, $f_1(0)$, 
is used, the effect of a right-handed currents $\epsilon_R^D$ manifest modifying the ``axial charge'':
\begin{eqnarray}
g_A^{\rm expt}=(1-2\eR^d)\,g_A~,\hspace{1cm}
g_1^{\rm expt}=(1-2\eR^s)\,g_1~.\label{eq:g1}
\end{eqnarray}
Thus, a bound on the right-handed current can be determined if any of the axial form factors is both measured and calculated in LQCD. 

The nonstandard coefficients $\epsilon_{S,P,T}^{D\ell}$ modify not only the total rate but also the kinematic distributions 
and polarization observables of the $\beta$ decays. Similarly to $K_{\ell3(\gamma)}$, the chiral suppression of (pseudo)scalar and 
tensor operators implies that only the muonic case presents a non-negligible linear dependence on the 
WC. For instance, in Ref.~\cite{Chang:2014iba} the following lepton universality ratio was studied with semileptonic hyperon decays:
\begin{align}
R_{B_1B_2}=\frac{\Gamma(B_1\to B_2\mu\nu)}{\Gamma(B_1\to B_2 e\nu)}= R_{B_1B_2}^{\rm SM}\,\frac{|\tilde V_{us}^{\mu}|^2}{|\tilde V_{us}^{e}|^2}(1+r_S\,\epsilon_S^{s\mu}+r_T\,\epsilon_T^{s\mu})
~,\label{eq:SHDNP}
\end{align}
\newpage
{\color{white} a}
\vspace{1.5cm}

\hspace{1.7cm}
\label{app:flowchart}
\begin{sideways}
\begin{tikzpicture}[scale=2, node distance = 2.7cm, auto]
    \node [redblock] (Kl3shape) {$K_{\ell 3}$ diff. distr.};
    \node [block, right of=Kl3shape,node distance=4cm] (shapeFF) {$I_e,I_\mu$,\\$\log C, B_T\epsilon_T^{s\mu}$};
    \node [wideblock, right of=shapeFF,node distance=5.8cm] (shapeWC) {$\epsilon_S^{s\mu}, \epsilon_T^{s\mu}$};
    \node [redblock, below of=Kl3shape] (Kl3rates) {$K_{\ell 3}$ rates};
    \node [block, right of=Kl3rates,node distance=4cm] (Kl3rates2) {$\tilde{V}_{us}^{e}f_+(0)$,\\$\tilde{V}_{us}^{\mu}f_+(0)$};
    \node [wideblock, right of=Kl3rates2,node distance=5.6cm] (Kl3ratesWC) {$\tilde{V}_{us}^{e},\tilde{V}_{us}^{\mu}$};
    \node [redblock, below of=Kl3rates,node distance=3.2cm] (Pl2) {$P_{\ell 2}$};
    \node [block, right of=Pl2,node distance=4cm] (Pl2bis) {$R_\pi, R_K$,\\$R_\mu,\Gamma_{K_{\mu2}}$};
    \node [wideblock, right of=Pl2bis,node distance=6cm] (Pl2WC) {
    ${\frac{|\tilde V^e_{ud}|^2}{|\tilde V^\mu_{ud}|^2}(1+\Delta^{\pi}_{e2/\mu2})}$\\\vspace{0.1cm}
    ${\frac{|\tilde V^e_{us}|^2}{|\tilde V^\mu_{us}|^2}(1+\Delta^K_{e2/\mu2})}$\\\vspace{0.1cm}
    $\frac{|\tilde V^e_{us}|^2}{|\tilde V^e_{ud}|^2}(1+\Delta^{K/\pi}_{\mu2})$\\\vspace{0.1cm}
    $|\tilde V_{us}^e|^2\,(1+\Delta^K_{\mu2})$};
    \node [redblock,below of=Pl2,node distance=3.2cm] (beta) {$\beta$ decays};
    \node [block, right of=beta,node distance=4cm] (init) {$\tilde{V}_{ud}^{e}, b_F$\\$R_{B_1B_2}$};
    \node [wideblock, right of=init,node distance=6cm] (betaWC) {$\tilde{V}_{ud}^{e}, \epsilon_S^{de}$ \\ $\epsilon_S^{s\mu}, \epsilon_T^{s\mu}$ };
    \node [block, below of=init,node distance=2.7cm] (gAtilde) {$g^{\rm expt}_A,\,g^{\rm expt}_1$};
    \node [wideblock, right of=gAtilde,node distance=6cm] (epsilonR) {$\epsilon_R^d, \epsilon_R^s$};
    \node [wideyellowblock, right of=Pl2WC,node distance=5cm] (preWC) {$\tilde{V}_{ud}^e$\\$\tilde{V}_{us}^e$\\ 
    $\Delta^s_L$\\
    $\Delta^{\pi}_{e2/\mu2} - \Delta_L^d$\\\vspace{0.1cm}
    $\Delta^K_{e2/\mu2}$\\$\Delta^{K/\pi}_{\mu2}$\\$ \Delta^K_{\mu2} $\\ $\epsilon_S^{d\mu}$\\$\epsilon_T^{d\mu}$\\$\epsilon_S^{de}$};
    \node [yellowblock, right of=preWC,node distance=4cm] (WC) {$\tilde{V}_{ud}^e$\\$\tilde{V}_{us}^e$\\ 
    $\Delta^s_L$\\
    $\Delta^d_{LP}$\\
    $\epsilon^{s\mu}_P$\\$\epsilon^{se}_P$\\ $\epsilon_S^{d\mu}$\\$\epsilon_T^{d\mu}$\\$\epsilon_S^{de}$\\$ \epsilon^{de}_P $\\
    $\epsilon_R^d$\\$\epsilon_R^s$};
   \node [right of=preWC,node distance=2.5cm] (prueba) {};
%
    \path [line] (beta) -- (init);
    \path [line] (init) -- node[decision answer] {\footnotesize{$g_{S},R_{S,T}$}} (betaWC);
    \path [line] (beta) -- (gAtilde);
    \path [line] (gAtilde) -- node[decision answer] {\footnotesize{$g_A,g_1$}} (epsilonR);
    \path [line] (Kl3shape) -- (shapeFF);
    \path [line] (shapeFF) -- node[decision answer] {\footnotesize{$\log C,B_T$}} (shapeWC);
    \path [dashedline] (shapeFF) -- (Kl3rates);
    \path [line] (Kl3rates) -- (Kl3rates2);
    \path [line] (Kl3rates2) -- node[decision answer] {\footnotesize{$f_+(0)$}} (Kl3ratesWC);
    \path [line] (Pl2) -- (Pl2bis);
    \path [line] (Pl2bis) -- node[decision answer] {\footnotesize{$f_K, \frac{f_K}{f_\pi}$}} (Pl2WC);
    \path [line] (shapeWC) -- (preWC);
    \path [line] (Kl3ratesWC) -- (preWC);
    \path [line] (Pl2WC) -- (preWC);
    \path [line] (betaWC) -- (preWC);
    \path [line] (preWC) -- (WC);
    \path [line] (epsilonR) -| (prueba);
\end{tikzpicture}
\\
\end{sideways}
\begin{sideways}
\end{sideways}
\newpage
\begin{table}[h!]
  \caption{Values of $r_{S,T}$ for the various semileptonic hyperon decays. \label{tab:senshyp}
  }
  \begin{center}
    \begin{tabular}{|ccccc|}    
      \hline 
            &  $\Lambda\to p$ & $\Sigma^{-}\!\to n$ & $\Xi^{0}\!\to\Sigma^{+}$ & $\Xi^{-}\!\to\Lambda$\tabularnewline
      \hline
      $r_S$&$1.60$&$4.1$&$0.56$&$3.7$\\
      $r_T$ & $5.2$ & $1.7$& $7.2$&$1.1$\\
      \hline 
    \end{tabular}
  \end{center}
\end{table}
\noindent where the coefficients $r_{S,T}$ embody the sensitivity of hyperon $\beta$ decays to $\epsilon_{S,T}^{s\mu}$. 
The lepton-flavor violating NP contributions to $|\tilde V_{us}^{\mu}|/|\tilde V_{us}^{e}|$ 
are encoded in $\Delta_L^s$, which can be extracted independently from $K_{\ell 3}$ decays, \textit{cf.} Eq.~(\ref{eq:Kl3LUrate}), so that measuring $R_{B_1B_2}$ in different 
channels leads to bounds on $\epsilon_{T,S}^{s\mu}$. 
It is interesting to note in Table~\ref{tab:senshyp}~\cite{Chang:2014iba} the complementarity between the different channels and the strong sensitivity of each decay to either 
$\epsilon_{S}^{s\mu}$ or $\epsilon_{T}^{s\mu}$. This contrasts with $K_{\mu 3}$ for the tensor WC where, as discussed above, the sensitivity of the total 
rates is a factor 10-20 smaller. 
New experiments and further analyses are necessary to exploit this interesting interplay between kaon and hyperon decays.

\section{Phenomenology}

In the previous page we present a flowchart summarizing how the processes described in the last section give access to different NP combinations 
of Wilson coefficients in a global (linearized) EFT analysis of NP in $D\to u\ell\nu$ transitions ($D=d,s$, $\ell=e,\mu$), 
along with the different experimental and theoretical inputs that are needed. A detailed 
explanation of the values employed in our numerical analysis can be found in Ref.~\cite{Gonzalez-Alonso:2016etj}. 
Treating all errors as Gaussian, we perform a standard $\chi^2$ fit, keeping only linear terms in $v^2/\Lambda^2$ in the theoretical expressions. The results
are:
\bea
\left(
\begin{array}{c}
 \tilde{V}_{ud}^{e} \\
 \tilde{V}_{us}^{e} \\
 \Delta_L^s \\
 \Delta^d_{LP} \\
 \epsilon_P^{de} \\
 \epsilon_R^d\\
 \epsilon_P^{se} \\
 \epsilon_P^{s\mu} \\
 \epsilon_R^s \\
 \epsilon_S^{s\mu} \\
 \epsilon_T^{s\mu} \\
 \epsilon_S^{de} \\
\end{array}
\right)
=
\left(
\begin{array}{c}
 0.97451\pm 0.00038 \\
 0.22408\pm 0.00087 \\
 1.0\pm 2.5 \\
 1.9\pm 3.8 \\
 4.0\pm 7.8 \\
 -1.3\pm 1.7 \\
 -0.4\pm 2.1 \\
 -0.7\pm 4.3 \\
 0.1\pm 5.0 \\
 -3.9\pm 4.9 \\
 0.5\pm 5.2 \\
 1.4\pm 1.3 \\
\end{array}
\right)
\times 10^{\wedge}
\left(
\begin{array}{c}
 0 \\
 0 \\
 -3 \\
 -2 \\
 -6 \\
 -2 \\
 -5 \\
 -3 \\
 -2 \\
 -4 \\
 -3 \\
 -3 \\
\end{array}
\right)~,
\eea
in the $\overline{MS}$ scheme at $\mu=2$ GeV. Let us remind the reader that $\Delta_L^s = \epsilon_{L}^{s\mu}-\epsilon_{L}^{se} $ 
and $\Delta^d_{LP}=  \eL^{de}-\eL^{d\mu}+24\eP^{d\mu}$.
The correlation matrix is given by:
\bea
\rho=
\left(
\scriptsize{
\begin{array}{cccccccccccc}
 1. & 0. & 0. & 0.01 & 0.01 & 0. & 0. & 0. & 0. & 0. & 0. & 0.82 \\
 - & 1. & -0.16 & 0. & 0. & 0. & 0.04 & 0.04 & 0. & -0.26 & 0. & 0. \\
 - & - & 1. & 0. & 0. & 0. & -0.01 & 0.02 & 0. & 0. & 0.46 & 0. \\
 - & - & - & 1. & 0.9995 & -0.87 & 0.09 & 0.09 & 0. & 0.04 & 0. & 0.01 \\
 - & - & - & - & 1. & -0.87 & 0.09 & 0.09 & 0. & 0.04 & 0. & 0.01 \\
 - & - & - & - & - & 1. & 0. & 0. & 0. & 0. & 0. & 0. \\
 - & - & - & - & - & - & 1. & 0.9995 & -0.98 & -0.01 & 0. & 0. \\
 - & - & - & - & - & - & - & 1. & -0.98 & -0.01 & 0.01 & 0. \\
 - & - & - & - & - & - & - & - & 1. & 0. & 0. & 0. \\
 - & - & - & - & - & - & - & - & - & 1. & 0. & 0. \\
 - & - & - & - & - & - & - & - & - & - & 1. & 0. \\
 - & - & - & - & - & - & - & - & - & - & - & 1. \\
\end{array}
}
\right)~.
\eea

Exploiting now the unitarity of the CKM matrix we can replace $\tilde{V}^e_{us}$ by
\bea
\Delta_{\rm CKM} &=& 1.9 (\eL^{de}+\eR^{d})+0.1(\eL^{se}+\eR^{s})-2\frac{\delta G_F}{G_F} =
 -(1.2\pm 8.4)\times 10^{-4}~,\nonumber\\
 \rho_{2i} &=&
 \left(
\scriptsize{
\begin{array}{cccccccccccc}
 0.88	& 1. & -0.07 & 0.01 & 0.01 & 0. & 0.02 & 0.02 & 0. & -0.12 & 0. & 0.73
\end{array}
}
\right)~,
\eea

We observe good agreement with the SM, with marginalized limits varying from the $10^{-5}$ level for the pseudoscalar couplings
in the electronic channel (due to the chiral enhancement) to the per-cent level for the right-handed couplings 
(due to the limited lattice precision in the axial-vector form factors). 

The combinations of WC $\{\Delta^d_{LP},\,\epsilon_P^{de}\}$ and 
$\{\epsilon_P^{se},\,\epsilon_P^{s\mu} \}$ are highly correlated, which simply reflects the fact that the specific combination 
of them that appears in $R_\pi$ and $R_K$ respectively is much more constrained than the individual WC. This is illustrated by the 
limits obtained when the only non-zero NP couplings are the pseudoscalar couplings in the electronic channel:
\bea
 \epsilon_P^{de} = (0.7\pm 2.5)\times 10^{-7}~,\\
 \epsilon_P^{se} = -(4.5\pm 3.7)\times 10^{-7}~.
\eea
Such strong bounds can be the result of a very high NP scale, $\Lambda \sim v / \sqrt{\epsilon} \sim {\cal O}(500)$ TeV, or a non-trivial structure in lepton-flavor 
space, such as $\eP^{D\ell}\sim m_\ell\,\eP^{D}$~\cite{Antonelli:2008jg,Alonso:2015sja}. 

\begin{figure}[h!]
\centerline{\includegraphics[width=0.5\textwidth]{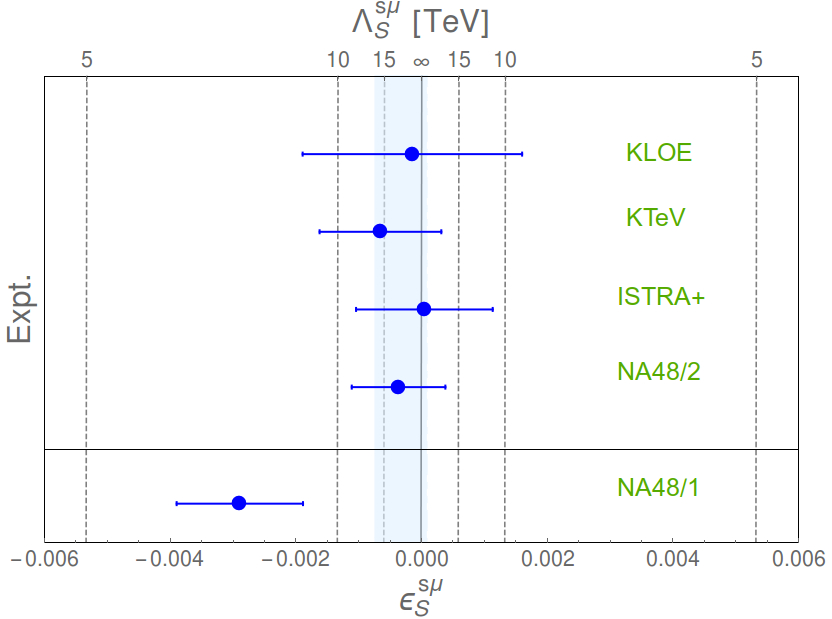}}
\caption{90\% C.L. constraints on the scalar WC $\epsilon_S^{s\mu}$ obtained from the different measurements of 
$\log C$~\cite{Yushchenko:2003xz,Alexopoulos:2004sy,Ambrosino:2007ad,Lai:2007dx,Abouzaid:2009ry,Antonelli:2010yf,Moulson:2014cra} and 
the theoretical determination using the CTT and the FLAG averages. We also plot the NA48/1 measurement, which we do not include in the average (blue band)~\cite{Antonelli:2010yf,Moulson:2014cra}. 
The NP effective scales are related to the WC as $\Lambda_S^{s\mu} \approx (V_{us} \epsilon_S^{s\mu})^{-1/2}v$. 
\label{fig:epsS}}   
\end{figure}

\subsection{The CTT theorem and the bound on $\epsilon_{S}^{s\mu}$}

As discussed above, a determination of the WC $\epsilon_S^{s\mu}$ crucially depends on our ability to 
predict the $q^2$ dependence of $f_0(q^2)$ in QCD and to perform precise extractions of it from the $K_{\mu 3}$ Dalitz plot. Luckily, the CTT theorem, combined with a determination
of $f_+(0)$ and $f_{K}/f_{\pi}$ in the lattice, leads to a very precise prediction of $\log\,C$, \textit{cf.} Eq.~(\ref{eq:lnCtilde}).
Experimentally, this quantity is extracted using the dispersive representation of the scalar 
form factor~\cite{Bernard:2006gy}. In Fig.~\ref{fig:epsS} we show the bounds on $\epsilon_S^{s\mu}$ obtained from the different measurements of 
$\log C$~\cite{Yushchenko:2003xz,Alexopoulos:2004sy,Ambrosino:2007ad,Lai:2007dx,Abouzaid:2009ry,Antonelli:2010yf,Moulson:2014cra} and 
the theoretical determination using the CTT and the FLAG averages~\cite{Aoki:2013ldr}, which rules out NP scales of $\Lambda_S^{s\mu}\lesssim 10$~TeV.

\subsection{The $V_{ud}-V_{us}$ plane revamped}
\label{sec:VudVusplot}

\begin{figure}[h!]
\centerline{\includegraphics[width=0.5\textwidth]{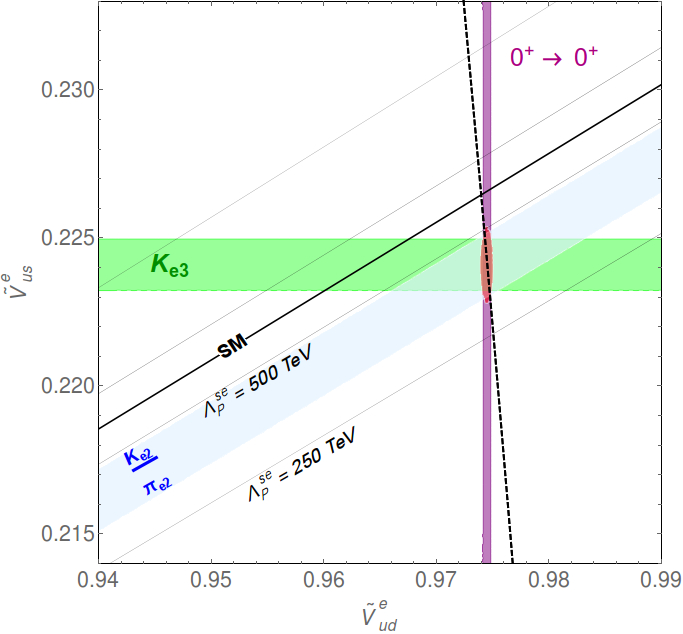}}
\caption{1$\sigma$ regions for $|\tilde V_{ud}^e|$ and $|\tilde V_{us}^e|$ from nuclear $\beta$ decays (vertical band), $K_{e3}$ (horizontal band), and their combination (red ellipse). We also plot the 1$\sigma$ region given by the ratio $\Gamma(K^\pm_{e2(\gamma)})/\Gamma(\pi^\pm_{e2(\gamma)})$ (diagonal band)
assuming a NP contribution $\Delta^{K/\pi}_{e2}\simeq0.02$, along with the lines corresponding to different NP effective scales $\Lambda_P^{se} \approx (V_{us} \epsilon_P^{se})^{-1/2}v$. 
The dashed black line shows the CKM unitarity constraint.
\label{fig:VudVus}}   
\end{figure}

If we look at the $|\tilde V^e_{ud}|-|\tilde V^e_{us}|$ plane, the NP can manifest either as a violation of CKM unitarity or as a ``misalignment'' of the
bound on $|\tilde V^e_{us}|/|\tilde V^e_{ud}|$ from $K_{e2}/\pi_{e2}$ with respect to the other two bounds from $\beta$ decays
($|\tilde V^e_{ud}|$) and $K_{e3}$ ($|\tilde V^e_{us}|$). The former case corresponds to the bound on $\Delta_{\rm CKM}$ obtained above, whereas
the latter probes the combination of Wilson Coefficients:
\bea
\Delta^{K/\pi}_e/2 = -2(\eR^s-\eR^d)- \frac{B_0}{m_e}\left(\eP^{se}-\eP^{de}\right)~,
\eea 
\textit{c.f.} Eqs.~(\ref{eq:epslKpi}-\ref{eq:deltarl1}). Hence, the right-handed and pseudoscalar contributions change the slope of the diagonal 
constraint obtained from $K_{e2}/\pi_{e2}$, as illustrated in Fig.~\ref{fig:VudVus}. We see that this precise test of the SM, obtained thanks 
to the small experimental and theoretical uncertainties achieved in these processes, currently allows one to probe ${\cal O}(100)$ TeV scales.

\subsection{Complementarity with collider searches}

If the new particles are too heavy to be produced on-shell at the LHC one can connect collider searches with 
low-energy processes in a model-independent way using the SMEFT~\cite{Bhattacharya:2011qm,Cirigliano:2012ab,Gonzalez-Alonso:2016etj}. 
The natural channel to study at the LHC is $pp\to \ell+ {\rm MET}+X$, since this process is sensitive at tree level to non-standard 
$\bar{u}s\to \ell\bar{\nu}$ partonic interactions:
\bea
\label{eq:sigmamt}
\sigma(m_T \!\!>\! \overline{m}_T) &=&
\sigma_W 
+ \sigma_S|\eS^{s\ell}|^2 +  \sigma_T |\eT^{s\ell}|^2 ~,
\eea  
where $\sigma_W$ represents the SM contribution, $\sigma_{S,T}(\overline{m}_T)$ are new functions which explicit form can be found in 
Ref.~\cite{Cirigliano:2012ab} and $\overline{m}_T$ is the chosen cut on the transverse mass.

\begin{figure}[h!]

\includegraphics[width=0.5\textwidth]{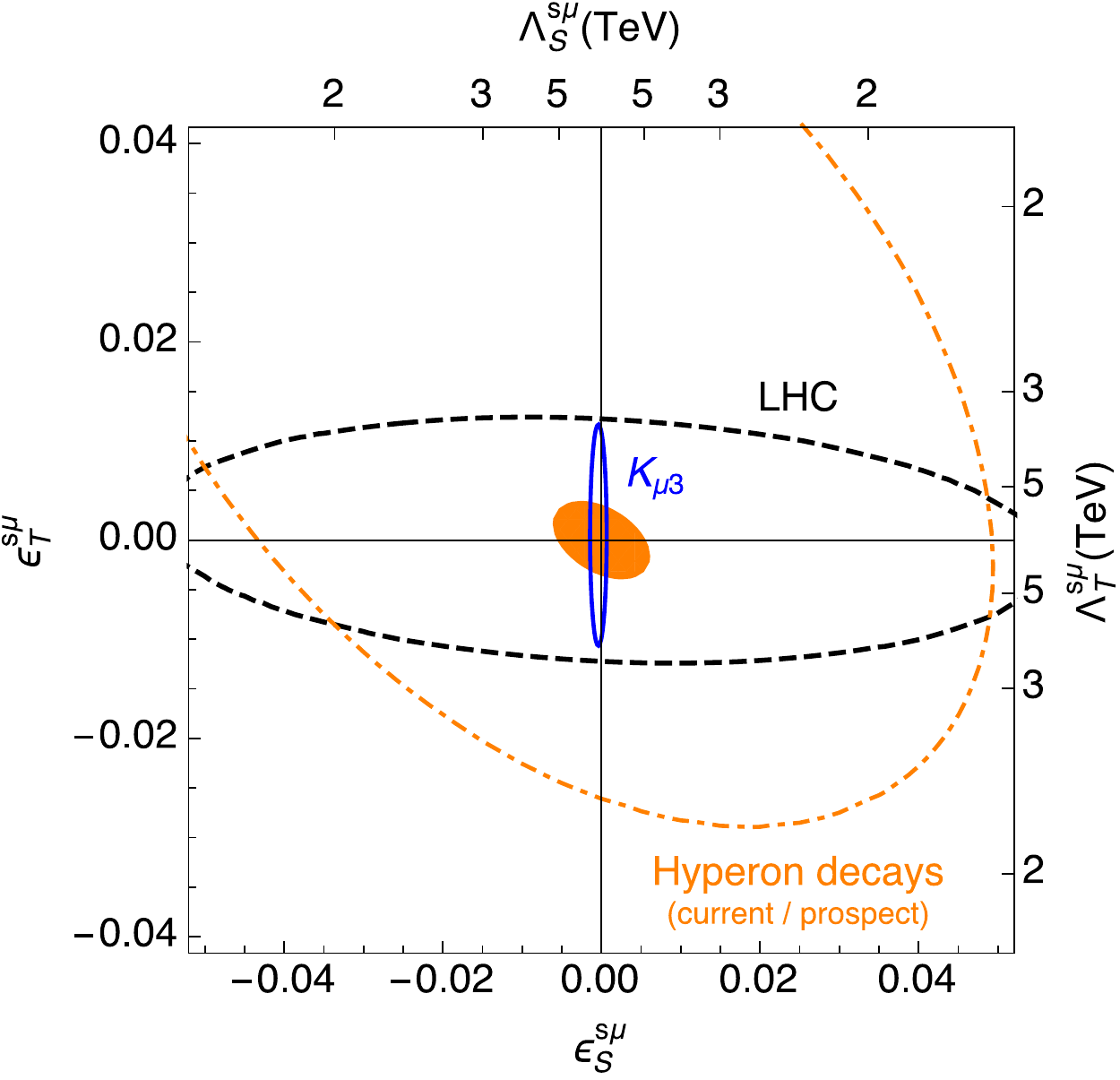}

\caption{90\% CL constraints on $\epsilon_{S,T}^{s\mu}$ at $\mu=2$ GeV from our global fit (blue solid ellipse), from the analysis 
of $pp\to \ell^\pm+{\rm MET}+X$ CMS data (black dashed ellipse) and from semileptonic hyperon decays (orange dot-dashed lines)~\cite{Chang:2014iba}. 
The filled orange ellipse shows the limits from hyperon decays assuming both experimental and theoretical errors reach the 1\% level.
Effective scales are defined by $\Lambda_i \approx (V_{us} \epsilon_i)^{-1/2} v$.
\label{fig:CPST}}   
\end{figure}

Using 20 fb$^{-1}$ of data recorded at $\sqrt{s}=$ 8 TeV by the CMS collaboration in the muonic channel  $pp\to \mu^\pm+{\rm MET}+X$\cite{Khachatryan:2014tva},  
one sees good agreement between data (3 events above $\overline{m}_T=1.5$~TeV) and SM ($2.35\pm 0.70$ events). This leads to the bound shown in Fig.~\ref{fig:CPST}, which is compared to the limit  
obtained from $K_{\mu3}$ in Ref.~\cite{Gonzalez-Alonso:2016etj} (from the Dalitz plot and using the CTT) and from hyperon $\beta$ decays in Ref.~\cite{Chang:2014iba} using the ratio
in Eq.~\eqref{eq:SHDNP}.~\footnote{The running of the LHC limits from 1 TeV to 2 GeV is performed using the one-loop QCD and EW renormalization group equations
~\cite{Gonzalez-Alonso:2016etj}} Finally, we also display 
a projection of the bounds from the hyperon decays in case the experimental data and the theoretical predictions both reached the 1\% level.
This illustrates the interesting complementarity between low-energy experiments and the LHC searches, where semileptonic decays are 
currently exploring new regions of the NP parameter space unaccessible to the LHC.

\section{Conclusions}
\label{sec:Conclusions}

In this note we succinctly describe a global model-independent approach to the analysis of NP in the $D\to u\ell\nu$ transitions ($D=d,s$; $\ell=e,\,\mu$).
We do not assume any flavor symmetry and we keep all possible NP operators at the same time.  We analyse (semi)leptonic kaon decays, pion decays and nuclear,
neutron and hyperon $\beta$ decays. The latter become necessary since one can not discriminate among all different possible NP effects using only pion and 
kaon decay observables. Besides providing a road map for future tests of the SM using all these processes, we provide numerical results of a fit using current 
experimental data and lattice QCD results. We find that these decays are sensitive to NP with typical scales of several TeV and we discussed complementarity 
with collider searches in the context of the SMEFT. 

More details can be found in the paper this note is based on~\cite{Gonzalez-Alonso:2016etj}.

\section*{Acknowledgements}

The authors would like to express a special thanks to the Mainz Institute for Theoretical Physics (MITP) for its hospitality and support and to
the organizers of the NA62 Physics Handbook workshop for giving us the chance to present
an early version of this work. M.G.-A. is grateful to the LABEX Lyon Institute of Origins (ANR-10-LABX-0066) 
of the Universit\'e de Lyon for its financial support within the program ANR-11-IDEX-0007 of the French government. 
JMC's work is funded by the People Programme (Marie Curie Actions) of the European Union's Seventh Framework Programme (FP7/2007-2013)
under REA grant agreement n PIOF-GA-2012-330458.

\bibliography{kaon.bib}
\end{document}